     \documentclass[aps,prd,preprint,superscriptaddress,showpacs,byrevtex,nofootinbib]{revtex4}
\usepackage{graphicx} 
\usepackage{dcolumn}  
\graphicspath{{ps}}

\renewcommand{\arraystretch}{1.1}

\newcommand{\mbc}{\ensuremath{M^{\prime}_{\rm bc}}}
\newcommand{\de}{\ensuremath{\Delta E}}
\newcommand{\Pep}{\ensuremath{e^{+}}}
\newcommand{\Pem}{\ensuremath{e^{-}}}
\newcommand{\APmuon}{\ensuremath{\mu^{+}}}
\newcommand{\Pmuon}{\ensuremath{\mu^{-}}}
\newcommand{\APlepton}{\ensuremath{\ell^{+}}}
\newcommand{\Plepton}{\ensuremath{\ell^{-}}}
\newcommand{\Pgpz}{\ensuremath{\pi^{0}}}
\newcommand{\Pgpp}{\ensuremath{\pi^{+}}}
\newcommand{\Pgpm}{\ensuremath{\pi^{-}}}
\newcommand{\PKst}{\ensuremath{K^{*+}}}
\newcommand{\PJpsi}{\ensuremath{J/\psi}}
\newcommand{\Pgyii}{\ensuremath{\psi(2S)}}
\newcommand{\Bz}{\ensuremath{B^{0}}}

\newcommand{\Bopp}{\ensuremath{B^{0}\to \psi(2S)\pi^{0}}}
\newcommand{\BpJKst}{\ensuremath{B^{+}\to J/\psi K^{*+}}}

\begin{document}

\begin{minipage}[]{0.6\columnwidth}
 \includegraphics[height=3.0cm,width=!]{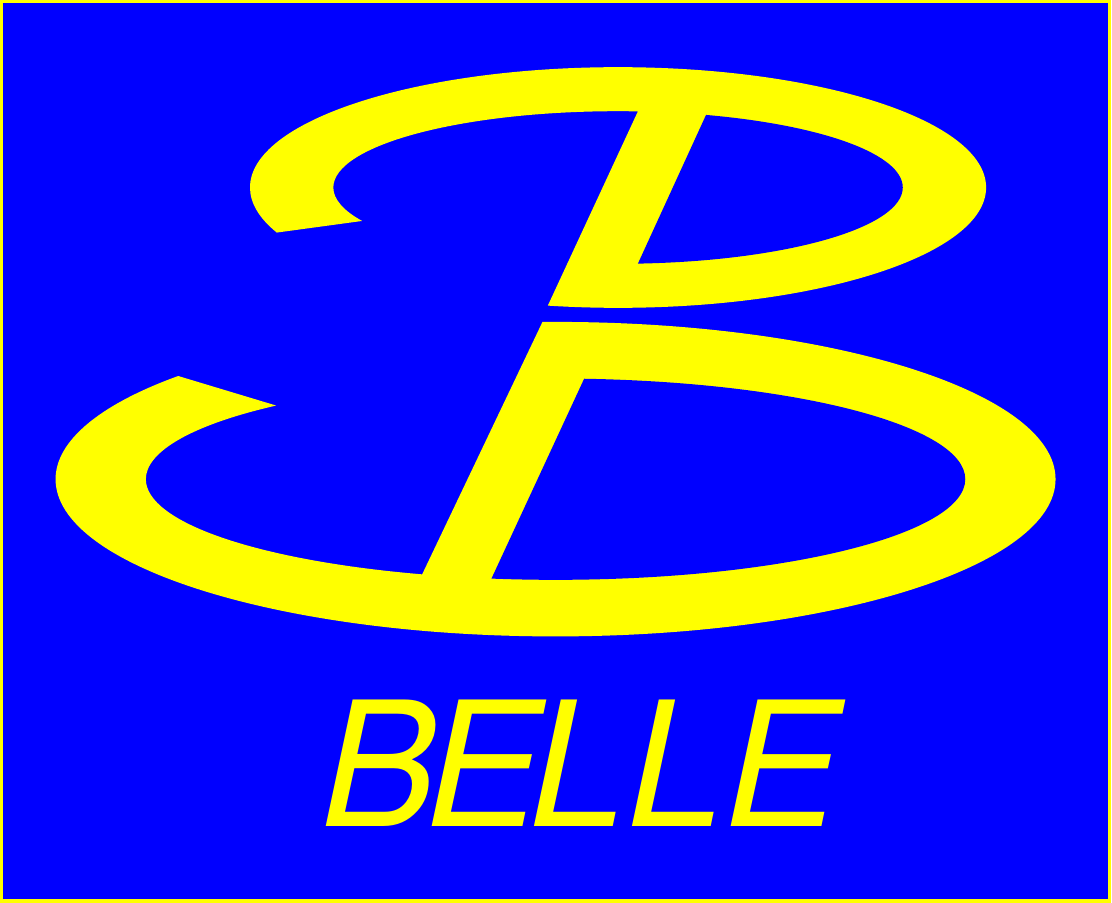} 
\end{minipage}
\begin{minipage}[]{0.4\columnwidth}
			\hbox{Belle Preprint {\it 15-20}}
			\hbox{KEK Preprint {\it 15-54}}
                        \hbox{}
\end{minipage}

\title{ \quad\\[1.0cm] First Observation of the Decay \Bopp}
\noaffiliation
\affiliation{Aligarh Muslim University, Aligarh 202002}
\affiliation{University of the Basque Country UPV/EHU, 48080 Bilbao}
\affiliation{Budker Institute of Nuclear Physics SB RAS, Novosibirsk 630090}
\affiliation{Faculty of Mathematics and Physics, Charles University, 121 16 Prague}
\affiliation{Chonnam National University, Kwangju 660-701}
\affiliation{University of Cincinnati, Cincinnati, Ohio 45221}
\affiliation{Deutsches Elektronen--Synchrotron, 22607 Hamburg}
\affiliation{University of Florida, Gainesville, Florida 32611}
\affiliation{Justus-Liebig-Universit\"at Gie\ss{}en, 35392 Gie\ss{}en}
\affiliation{Gifu University, Gifu 501-1193}
\affiliation{SOKENDAI (The Graduate University for Advanced Studies), Hayama 240-0193}
\affiliation{Hanyang University, Seoul 133-791}
\affiliation{University of Hawaii, Honolulu, Hawaii 96822}
\affiliation{High Energy Accelerator Research Organization (KEK), Tsukuba 305-0801}
\affiliation{IKERBASQUE, Basque Foundation for Science, 48013 Bilbao}
\affiliation{Indian Institute of Technology Bhubaneswar, Satya Nagar 751007}
\affiliation{Indian Institute of Technology Guwahati, Assam 781039}
\affiliation{Indian Institute of Technology Madras, Chennai 600036}
\affiliation{Indiana University, Bloomington, Indiana 47408}
\affiliation{Institute of High Energy Physics, Chinese Academy of Sciences, Beijing 100049}
\affiliation{Institute of High Energy Physics, Vienna 1050}
\affiliation{Institute for High Energy Physics, Protvino 142281}
\affiliation{INFN - Sezione di Torino, 10125 Torino}
\affiliation{Institute for Theoretical and Experimental Physics, Moscow 117218}
\affiliation{J. Stefan Institute, 1000 Ljubljana}
\affiliation{Kanagawa University, Yokohama 221-8686}
\affiliation{Institut f\"ur Experimentelle Kernphysik, Karlsruher Institut f\"ur Technologie, 76131 Karlsruhe}
\affiliation{Kennesaw State University, Kennesaw GA 30144}
\affiliation{King Abdulaziz City for Science and Technology, Riyadh 11442}
\affiliation{Korea Institute of Science and Technology Information, Daejeon 305-806}
\affiliation{Korea University, Seoul 136-713}
\affiliation{Kyungpook National University, Daegu 702-701}
\affiliation{\'Ecole Polytechnique F\'ed\'erale de Lausanne (EPFL), Lausanne 1015}
\affiliation{Faculty of Mathematics and Physics, University of Ljubljana, 1000 Ljubljana}
\affiliation{Ludwig Maximilians University, 80539 Munich}
\affiliation{Luther College, Decorah, Iowa 52101}
\affiliation{University of Maribor, 2000 Maribor}
\affiliation{Max-Planck-Institut f\"ur Physik, 80805 M\"unchen}
\affiliation{School of Physics, University of Melbourne, Victoria 3010}
\affiliation{Moscow Physical Engineering Institute, Moscow 115409}
\affiliation{Moscow Institute of Physics and Technology, Moscow Region 141700}
\affiliation{Graduate School of Science, Nagoya University, Nagoya 464-8602}
\affiliation{Kobayashi-Maskawa Institute, Nagoya University, Nagoya 464-8602}
\affiliation{Nara Women's University, Nara 630-8506}
\affiliation{National Central University, Chung-li 32054}
\affiliation{National United University, Miao Li 36003}
\affiliation{Department of Physics, National Taiwan University, Taipei 10617}
\affiliation{H. Niewodniczanski Institute of Nuclear Physics, Krakow 31-342}
\affiliation{Niigata University, Niigata 950-2181}
\affiliation{Novosibirsk State University, Novosibirsk 630090}
\affiliation{Osaka City University, Osaka 558-8585}
\affiliation{Pacific Northwest National Laboratory, Richland, Washington 99352}
\affiliation{Panjab University, Chandigarh 160014}
\affiliation{University of Pittsburgh, Pittsburgh, Pennsylvania 15260}
\affiliation{Punjab Agricultural University, Ludhiana 141004}
\affiliation{University of Science and Technology of China, Hefei 230026}
\affiliation{Seoul National University, Seoul 151-742}
\affiliation{Soongsil University, Seoul 156-743}
\affiliation{University of South Carolina, Columbia, South Carolina 29208}
\affiliation{Sungkyunkwan University, Suwon 440-746}
\affiliation{School of Physics, University of Sydney, NSW 2006}
\affiliation{Department of Physics, Faculty of Science, University of Tabuk, Tabuk 71451}
\affiliation{Tata Institute of Fundamental Research, Mumbai 400005}
\affiliation{Excellence Cluster Universe, Technische Universit\"at M\"unchen, 85748 Garching}
\affiliation{Department of Physics, Technische Universit\"at M\"unchen, 85748 Garching}
\affiliation{Toho University, Funabashi 274-8510}
\affiliation{Department of Physics, Tohoku University, Sendai 980-8578}
\affiliation{Earthquake Research Institute, University of Tokyo, Tokyo 113-0032}
\affiliation{Department of Physics, University of Tokyo, Tokyo 113-0033}
\affiliation{Tokyo Institute of Technology, Tokyo 152-8550}
\affiliation{Tokyo Metropolitan University, Tokyo 192-0397}
\affiliation{University of Torino, 10124 Torino}
\affiliation{Utkal University, Bhubaneswar 751004}
\affiliation{CNP, Virginia Polytechnic Institute and State University, Blacksburg, Virginia 24061}
\affiliation{Wayne State University, Detroit, Michigan 48202}
\affiliation{Yamagata University, Yamagata 990-8560}
\affiliation{Yonsei University, Seoul 120-749}
  \author{V.~Chobanova}\affiliation{Max-Planck-Institut f\"ur Physik, 80805 M\"unchen} 
  \author{J.~Dalseno}\affiliation{Max-Planck-Institut f\"ur Physik, 80805 M\"unchen}\affiliation{Excellence Cluster Universe, Technische Universit\"at M\"unchen, 85748 Garching} 
  \author{C.~Kiesling}\affiliation{Max-Planck-Institut f\"ur Physik, 80805 M\"unchen} 
  \author{A.~Abdesselam}\affiliation{Department of Physics, Faculty of Science, University of Tabuk, Tabuk 71451} 
  \author{I.~Adachi}\affiliation{High Energy Accelerator Research Organization (KEK), Tsukuba 305-0801}\affiliation{SOKENDAI (The Graduate University for Advanced Studies), Hayama 240-0193} 
  \author{H.~Aihara}\affiliation{Department of Physics, University of Tokyo, Tokyo 113-0033} 
  \author{D.~M.~Asner}\affiliation{Pacific Northwest National Laboratory, Richland, Washington 99352} 
  \author{T.~Aushev}\affiliation{Moscow Institute of Physics and Technology, Moscow Region 141700}\affiliation{Institute for Theoretical and Experimental Physics, Moscow 117218} 
  \author{R.~Ayad}\affiliation{Department of Physics, Faculty of Science, University of Tabuk, Tabuk 71451} 
  \author{V.~Babu}\affiliation{Tata Institute of Fundamental Research, Mumbai 400005} 
% ...
\collaboration{The Belle Collaboration}

\noaffiliation

\begin{abstract}
We report a measurement of the \Bopp\ branching fraction based on the full $\Upsilon(4S)$ data set of $772 \times 10^6$ $B\bar{B}$ pairs  
collected by the Belle detector at the KEKB asymmetric-energy $e^+ e^-$ collider. We obtain ${\cal B}(\Bopp) = (1.17\pm0.17\text{(stat)}\pm0.08\text{(syst)})\times10^{-5}$.
The result has a significance of 7.2 standard deviations and is the first observation of the decay \Bopp. 
\end{abstract}


\maketitle

\tighten

{\renewcommand{\thefootnote}{\fnsymbol{footnote}}}
\setcounter{footnote}{0}

Violation of the combined charge--parity symmetry ($CP$ violation) in the Standard Model (SM) arises from a single irreducible phase in the Cabibbo--Kobayashi--Maskawa 
(CKM) quark-mixing matrix~\cite{C,KM}. A primary objective of the Belle experiment is to overconstrain the unitarity triangle of the CKM matrix related to $B_{u,d}$ decays. 
This permits a precision test of the CKM mechanism for $CP$ violation as well as the search for effects beyond the SM. Mixing-induced $CP$ violation in the $B$ sector has 
been clearly established by the Belle~\cite{phi1_Belle} and BaBar~\cite{phi1_BaBar} collaborations in the $b \to c \bar c s$-induced decays $B^0 \to (c \bar c)^{0} K^{0}$.

While these decays allow access to the $CP$ violating angle $\phi_{1} \equiv \arg(-V_{cd}V^{*}_{cb})/(V_{td}V^{*}_{tb})$ at first order (tree), its value is prone to 
distortion from suppressed higher-order loop-induced (penguin) amplitudes containing different weak phases. Applying SU(3) symmetry arguments, the related 
$b \to c \bar c d$-induced channels $B^0 \to (c \bar c)^{0} \pi^{0}$ can be used to quantify the shift in $\phi_{1}$ caused by these loop contributions~\cite{psi2spi0}. 
Thus, this $b\to c\bar{c}d$ decay is a promising place to search for new physics effects~\cite{GrossmanWorah}. This paper establishes the $B^0 \to \psi(2S) \pi^{0}$ channel, 
which may be used to constrain the penguin contamination in $\Bz \to \psi(2S) K^{0}$ in a future measurement of its time-dependent $CP$ asymmetry.

The result presented in this paper is based on the final $\Upsilon(4S)$ data sample, containing $772 \times 10^{6}$ $B\bar{B}$ pairs collected with the Belle detector at 
the KEKB asymmetric-energy $e^{+}e^{-}$ ($3.5$ on $8~{\rm GeV}$) collider~\cite{KEKB}. At the $\Upsilon(4S)$ resonance, corresponding to a center-of-mass energy 
$\sqrt{s} = 10.58$~GeV, the $B\bar{B}$ pairs are produced with a Lorentz boost $\beta\gamma =0.425$ nearly along the $+z$ direction, which is opposite the positron beam 
direction.

The Belle detector is a large-solid-angle magnetic spectrometer that consists of a silicon vertex detector (SVD), a 50-layer central drift chamber (CDC), an array of
aerogel threshold Cherenkov counters (ACC), a barrel-like arrangement of time-of-flight scintillation counters (TOF), and an electromagnetic calorimeter (ECL) comprising 
CsI(Tl) crystals located inside 
a superconducting solenoid coil that provides a 1.5~T magnetic field.  An iron flux-return yoke located outside of the coil is instrumented to detect $K_L^0$ mesons and to identify
muons (KLM).  The detector is described in detail elsewhere~\cite{Belle}. Two inner detector configurations were used: A 2.0 cm radius beampipe and a three-layer silicon vertex 
detector (SVD1) were used for the first sample of $152 \times 10^6$ $B\bar{B}$ pairs, while a 1.5 cm radius beampipe, a four-layer silicon vertex detector (SVD2), and a small-cell 
inner drift chamber were used to record  the remaining $620 \times 10^6$ $B\bar{B}$ pairs~\cite{svd2}. Simulated $B$ decay Monte Carlo (MC) events are generated by 
EvtGen~\cite{evtgen}, in which final-state radiation is described with PHOTOS~\cite{photos}. We use the GEANT3~\cite{GEANT} toolkit to model the interaction of the generated 
particles with the detector and its response in order to determine the detector acceptance.
 
We reconstruct $\psi(2S)$ candidates in the $\APlepton\Plepton$ decay channels ($\ell = e, \mu$), referred to as leptonic hereinafter, and the $J/\psi \pi^{+} \pi^{-}$ decay 
channel, referred to as hadronic. All charged tracks are identified using a loose requirement on the distance of closest approach with respect to the interaction
point along the beam direction of under $5.0\;\rm  cm$ and in the transverse plane of under $1.5\;\rm cm$.
The $J/\psi$ candidates are  reconstructed from $\APlepton\Plepton$ pairs. Electron tracks are identified by a combination of $dE/dx$ in the CDC, shower shape and position in the ECL, 
light yield in the ACC, and $E/p$, where $E$ is the energy deposited in the ECL and $p$ is the momentum measured by the SVD and the CDC. To account for radiative 
energy losses in the $e^{+} e^{-}$ decays, we include the bremsstrahlung photons ($\gamma$) that are in a cone with an opening angle of $50\;\rm mrad$ around the 
$e^{+}$ ($e^{-}$) tracks [so that the reconstructed $J/\psi$ or \Pgyii\ candidate is denoted as $e^{+} e^{-} (\gamma)$]. For muon tracks, the identification is 
based on track penetration depth and hit scatter in the KLM.

We impose asymmetric requirements on the \PJpsi\ and \Pgyii\ masses due to energy leakage in the ECL and bremsstrahlung.
The invariant masses of the $J/\psi$ candidates must fulfill $M_{e^{+}e^{-}(\gamma)}-m_{J/\psi} \in (-0.150,+0.036) {\;\rm GeV}/c^{2}$ or 
$M_{\mu^{+}\mu^{-}} - m_{J/\psi} \in (-0.060,+0.036) {\;\rm GeV}/c^{2}$, where $m_{J/\psi}$ denotes 
the world-average $J/\psi$ mass~\cite{PDG}, and $M_{e^{+}e^{-}(\gamma)}$ and $M_{\mu^{+}\mu^{-}}$ are the reconstructed invariant masses of the $e^{+} e^{-} (\gamma)$ and 
$\mu^{+} \mu^{-}$ candidates, respectively. For the $\psi(2S)$, the invariant masses must fulfill
$ M_{e^{+}e^{-}(\gamma)} - m_{\psi(2S)} \in (-0.150,+0.036) {\;\rm GeV}/c^{2}$ or $M_{\mu^{+}\mu^{-}} - m_{\psi(2S)} \in (-0.060,+0.036) {\;\rm GeV}/c^{2}$, where 
$m_{\psi(2S)}$ denotes the world-average $\psi(2S)$ mass~\cite{PDG}. For the $\psi(2S) \to J/\psi \pi^{+} \pi^{-}$ candidates, 
$\Delta M \equiv M_{\APlepton\Plepton(\gamma)\pi^{+} \pi^{-}} - M_{\APlepton\Plepton(\gamma)}$ must fulfill $\Delta M \in (0.580,0.600) {\;\rm GeV}/c^{2}$. To reduce background particle 
combinations in this channel, we select $\Pgpp\Pgpm$ pairs with an invariant mass above a loose threshold of $400{\;\rm MeV}/c^{2}$. 
Using information obtained from the CDC, ACC, and TOF, these pion candidates are also required to be inconsistent with the kaon mass hypothesis. This requirement retains 99.8\% of the
pion candidates, while 5\% of kaons are falsely identified as pions. To improve the $B$ meson mass resolution,
we apply a vertex- and mass-constrained kinematic fit to the \PJpsi\ and \Pgyii\ candidates. We assign each candidate its nominal mass and require that its charged daughters
originate from the same vertex.

Photons are identified as isolated ECL clusters that are not matched to any charged particle track. To suppress combinatorial background, the photons are required to have  
energies above $50 \;\rm MeV$ if in the ECL barrel or above $100 \;\rm MeV$ if in the ECL endcaps, where the barrel region covers the polar angle range 
$32^{\circ} < \theta < 130^{\circ}$ and the endcap regions cover the polar angle ranges $12^{\circ} < \theta < 32^{\circ}$ and 
$130^{\circ} < \theta < 157^{\circ}$. Two $\gamma$ candidates are combined to form a $\pi^{0}$ candidate that must satisfy 
$ M_{\gamma\gamma} - m_{\pi^{0}} \in (-17,15) {\;\rm MeV}/c^{2}$, where $m_{\pi^{0}}$ is the world-average mass of the $\pi^{0}$~\cite{PDG}. This corresponds to about
three times the experimental resolution. The four-momenta of retained candidates are then adjusted in a mass-constrained fit wherein the parent mass is constrained to $m_{\pi^{0}}$.

We combine the $\psi(2S)$ and $\pi^{0}$ to form a neutral $B$ meson. The $B$ candidates are identified using two kinematic variables: a modified beam-energy-constrained mass,
\begin{widetext}
 \begin{equation}
  \mbc \equiv \sqrt{\left(E_{\rm beam}\right)^2-\left|\vec{p}_{\psi(2S)} + 
  \sqrt{\left(E_{\rm beam}-E_{\psi(2S)}\right)^2 - m_{\pi^{0}}^{2}} \frac{\vec{p}_{\pi^{0}}}
   {{|\vec{p}_{\pi^{0}}|} }\right|^{2}},
 \end{equation}
\end{widetext}
and the energy difference $\Delta E \equiv E_{B} - E_{\rm beam}$, 
where $\vec{p}$ denotes 3-momentum and $E_{\rm beam}$ the beam energy, all evaluated in the $\Upsilon(4S)$ center-of-mass system. This definition of 
\mbc\ is preferred over the standard form used at the $B$ factories as it exhibits a lower correlation with $\Delta E$ when $\pi^{0}$ is present in the final state. 

A significant background arises from $e^{+}e^{-} \to q \bar q$ $(q = u, d, s, c)$ continuum events. To suppress it, we construct the ratio of  second- 
to zeroth-order Fox--Wolfram moments~\cite{FoxWolfram}, $R_{2} = H_{2}/H_{0}$, which ranges between zero (spherical) and one (jet-like). 
A loose requirement of less than 0.5 is applied. This removes around $50\%$ of all continuum background with a negligible loss of
signal efficiency.

On average, 1.13 $B^{0}$ candidates are reconstructed per event and 11.6\% of all events have more than one candidate. In a multi-candidate event, we choose the $B^{0}$ with the lowest 
$\chi_{\rm mass}^{2} \equiv (M_{\rm Rec} - m)^2/\sigma^2_{\rm Rec}$ per daughter particle with a reconstructed mass $M_{\rm Rec}$, a nominal mass $m$ 
and a mass resolution $\sigma_{\rm Rec}$. 
For the leptonic channels, $\chi_{\rm mass}^{2} \equiv (\chi^2_{\psi(2S)} + \chi^2_{\pi^0})/2$.
For the hadronic channels, $\chi_{\rm mass}^{2} \equiv (\chi^2_{J/\psi} + \chi^2_{\Delta m} + \chi^2_{\pi^0})/3$, where $\chi_{\Delta m}^2$ is defined similarly except that the 
reconstructed and nominal mass differences between $\psi(2S)$ and $J/\psi$ are used in place of $M_{\rm Rec}$ and $m$, respectively. According to MC simulation, this procedure 
has a 75\% success rate when more than one $B$ candidate is reconstructed and the correct $B$ is in the list. After this best-candidate selection, 
the detection efficiency, including a correction for the difference between data and MC in the particle identification and including the daughter branching 
fraction uncertainties and the selection criteria uncertainties, is $(0.43 \pm 0.02)\%$ for the leptonic channels and $(0.52 \pm 0.02)\%$ for the hadronic. 
Approximately 0.5\% (10\%) of the signal candidates are misreconstructed in the leptonic (hadronic) channels.

The \Bopp\ branching fraction, ${\cal B}(\Bopp)$, is extracted from an unbinned extended maximum likelihood fit to \mbc\ and \de.
The following categories are considered in the event model: correctly-reconstructed signal, misreconstructed signal, other $b\to (c \bar{c})q$ transitions, 
and combinatorial background. Unless otherwise stated, the probability density function (PDF) is the 
product of PDFs for each observable, ${\cal P}^{m}_{c}(\mbc,\de) \equiv {\cal P}^{m}_{c}(\mbc)\times{\cal P}^{m}_{c}(\de)$, in each \Pgyii\ decay mode, $m$, and in 
each category, $c$.

We study the distributions of both signal components -- correctly reconstructed and misreconstructed -- using an MC sample that contains only \Bopp\ events.
We define a correctly-reconstructed event as one in which all charged tracks are correctly associated with the signal $B$ meson.
For such events, we find the distributions of the fit observables in the $\Pgyii\to\Pep\Pem$ and $\Pgyii\to\PJpsi[\Pep\Pem]\Pgpp\Pgpm$ decay 
channels to be similar. The distributions in the $\Pgyii\to\APmuon\Pmuon$ and $\Pgyii\to\PJpsi[\APmuon\Pmuon]\Pgpp\Pgpm$ decay modes are also alike. Thus, we divide 
the signal MC into an electron and a muon component and model these separately. The \mbc\ PDF for both modes consists of a Crystal Ball (CB) function~\cite{CB}, $\mathcal{C}$, 
combined with an ARGUS distribution~\cite{ARGUS}, $\mathcal{A}$, which additionally accounts for the tail towards 
lower \mbc\ values due to the photon and electron energy leakage in the ECL. Due to a correlation between \mbc\ and \de, we parametrize the \mbc\ PDF in terms of \de,
\begin{eqnarray}
  {\cal P}^{m}_{\rm Sig}(\mbc|\de)&\!\!\!\equiv\!\!\!& 
  (f^{m}+\rho_{1}^{m}\de^{2})\mathcal{C}(\mbc; \alpha_{\mbc}^{m}, \, n_{\mbc}^{m},\nonumber\\ 
  &&  \, \mu_{\mbc}^{m}+\mu_{\mbc}^{\rm CF}, \, \sigma^{m}_{\mbc}\sigma_{\mbc}^{\rm CF}+\rho_{2}^{m}g^{m}(\de)) \nonumber \\
  &&+ (1-[f^{m}+\rho_{1}^{m}\de^{2}]) \mathcal{A}(\mbc; a^{m}),
\end{eqnarray}
where $\alpha_{\mbc}^{m}$, $n_{\mbc}^{m}$, $\mu_{\mbc}^{m}$, $\sigma^{m}_{\mbc}$ and $a^{m}$ are parameters obtained from MC, while $\mu_{\mbc}^{\rm CF}$ and 
$\sigma_{\mbc}^{\rm CF}$ are correction factors obtained from a \BpJKst\ control sample; $\rho_{1}^{m}$ and $\rho_{2}^{m}$ are correlation factors and $g^{m}(\de)$ 
are functions in \de\ determined from MC: $g^{\Pep\Pem} = \de^{2}$ for the electron component and $g^{\APmuon\Pmuon} = |\de|$ for the muon 
component. For both types of correctly reconstructed signal events, the \de\ PDF is the combination of a CB distribution and a sum of
Chebyshev polynomials up to the first order,
\begin{eqnarray}
  {\cal P}^{m}_{\rm Sig}(\de)&\equiv& f^{m}\mathcal{C}(\de; \alpha_{\de}^{m}, \, n_{\de}^{m}, \, \mu_{\de}^{m}+\mu_{\de}^{\rm CF}, \, \sigma_{\de}^{m}\sigma_{\de}^{\rm CF}) \nonumber \\
  &&+ (1-f^{m}) (1 + c^{m} \de ),
\end{eqnarray}
where $\alpha^{m}_{\de}$, $n_{\de}^{m}$, $\mu_{\de}^{m}$, $\sigma_{\de}^{m}$ and $c^{m}$ are obtained from MC, while $\mu_{\de}^{\rm CF}$ and $\sigma_{\de}^{\rm CF}$ are correction factors 
obtained from the control sample.

We omit the misreconstructed signal component in the leptonic decay modes due to its insignificant contribution. Each of the two hadronic modes is modeled with a separate 
two-dimensional histogram in \mbc--\de. 

The major background contribution originates from $b\to (c\bar{c})q$ decays other than the signal. We study this component from an MC sample containing all known $b\to (c\bar{c})q$ 
decays.
Since the two leptonic channels have similar distributions, as do the two hadronic channels, we divide the $b\to (c\bar{c})q$ background events into a leptonic and a hadronic 
subsample. We model each of these with a two-dimensional \mbc--\de\ histogram.

The rest of the background events are a mixture of $\Pep\Pem\to q\bar{q}$ ($q=u,d,s,c$) processes and $B$ meson decays into open charm and charmless final states.
We refer to these as combinatorial background. We study their distributions from $\Upsilon(4S)$ data in the dilepton and $\Delta M$ sidebands.
The \PJpsi\ sideband is defined as $M_{\APlepton\Plepton}\in(2.60,2.80)\cup(3.20,3.40){\;\rm GeV}/c^{2}$, the \Pgyii\ sideband as 
$M_{\APlepton\Plepton}\in(3.45,3.53)\cup(3.80,3.90){\;\rm GeV}/c^{2}$, and the $\Delta M$ sideband as $\Delta M \in (0.49,0.53)\cup(0.64,0.68){\;\rm GeV}/c^{2}$.

In all sidebands, the \mbc\ PDF is an ARGUS distribution.
In the leptonic sidebands, we model the \de\ combinatorial background distribution with a sum of Chebyshev polynomials up to the first order. 
The combinatorial \de\ PDF in the $\Delta M$ sideband is a sum of Chebyshev polynomials up to the second order. We verify that the models in the lower and
upper sidebands are in agreement and thus the combined model provides a reliable description of the events in the signal region.

The total extended likelihood is given by
\begin{equation}
  {\cal L} \equiv \prod_{m} \frac{e^{-\sum_{c}N^{m}_{c}}}{N^{m}!} \prod^{N^{m}}_{i=1} 
  \sum_{c}N_{c} {\cal P}^{m}_{c}(M_{\rm bc}^{\prime \;i}, \de^{i}),
\end{equation}
where $i$ indexes the events, $c$ the categories and $m$ the decay modes.

The \Bopp\ branching fraction is a free parameter in the fit to the data and is obtained by transforming the signal yields according to
\begin{equation}
 N^{m}_{\rm Sig} = {\cal B}(\Bopp)N_{B\bar{B}} \epsilon_{\rm Sig}^{m},
\end{equation}
where $N_{B\bar{B}}$ is the number of $B\bar{B}$ pairs collected by the Belle detector and $\epsilon_{\rm Sig}^{m}$ is the detection efficiency, including daughter 
branching fractions for each subcategory. 
The misreconstructed-signal yields are fixed from MC relative to the two hadronic-mode signal yields. Only the
muonic hadronic mode's yield is free in the $c\bar{c}$ background category, while the yields of the three remaining decay modes are fixed 
from MC relative to it. The four combinatorial-background yields are free. 

We study the fit performance using pseudo-experiments in a linearity test covering the region of the expected branching fraction. There is no bias in experiments
where the events are generated according to the total PDF. However, a bias at the level of 10\% of the statistical error tending towards higher values is observed 
in experiments generated by selecting random events from the MC samples that have passed the full selection. This indicates that the bias is not due to a low
signal yield but rather to imperfections in the modeling of correlations. We apply a fit correction of the full bias and consider half the correction as a 
systematic uncertainty.

The contribution of peaking background that originates from decays to the same final state as the signal is studied in the \PJpsi, \Pgyii\ and $\Delta m$ sidebands.
We define the combinatorial background as non-peaking in \mbc\ and \de, while we assume that a potential peaking background has the same shape as the correctly 
reconstructed signal. Using the combinatorial background and the signal PDFs in a common fit to the sidebands, we extract two yields: one for the combinatorial background and the other 
for the peaking background. The peaking-background yield is consistent with zero for all modes except for the muonic signal mode in the $\Delta M$ sideband, where it has a 
statistical significance of $3.7 \sigma$. We extrapolate the expected peaking background yield into the signal region and subtract the obtained value from the 
signal yield obtained from the data.

We determine the \mbc\ and \de\ signal model correction factors from a control sample with a similar decay topology, \BpJKst, where the $\PKst$ candidates are 
reconstructed from a $K^{+}$ and a \Pgpz\ candidate. To ensure a high momentum of the \Pgpz, replicating the kinematic conditions of \Bopp,
we require the angle between the \Pgpz\ momentum vector and the vector opposite the $B$ flight direction in the $\PKst$ rest frame to be smaller than $1.5\;{\rm rad}$. 
For the \PJpsi\ and \Pgpz\ candidates, we use the same selection criteria as for the \Bopp\ mode. Only $\PKst$ candidates fulfilling 
$ M_{K^{+}\Pgpz}\in (0.793,0.990)\;{\rm GeV}/c^2$ are retained. Using a model similar to \Bopp\ for the control sample, we obtain a \BpJKst\ 
signal yield of $3681\pm71$ events and the signal correction factors from the fit to the data.

\begin{figure}
 \centering
 \includegraphics[width=0.49\columnwidth]{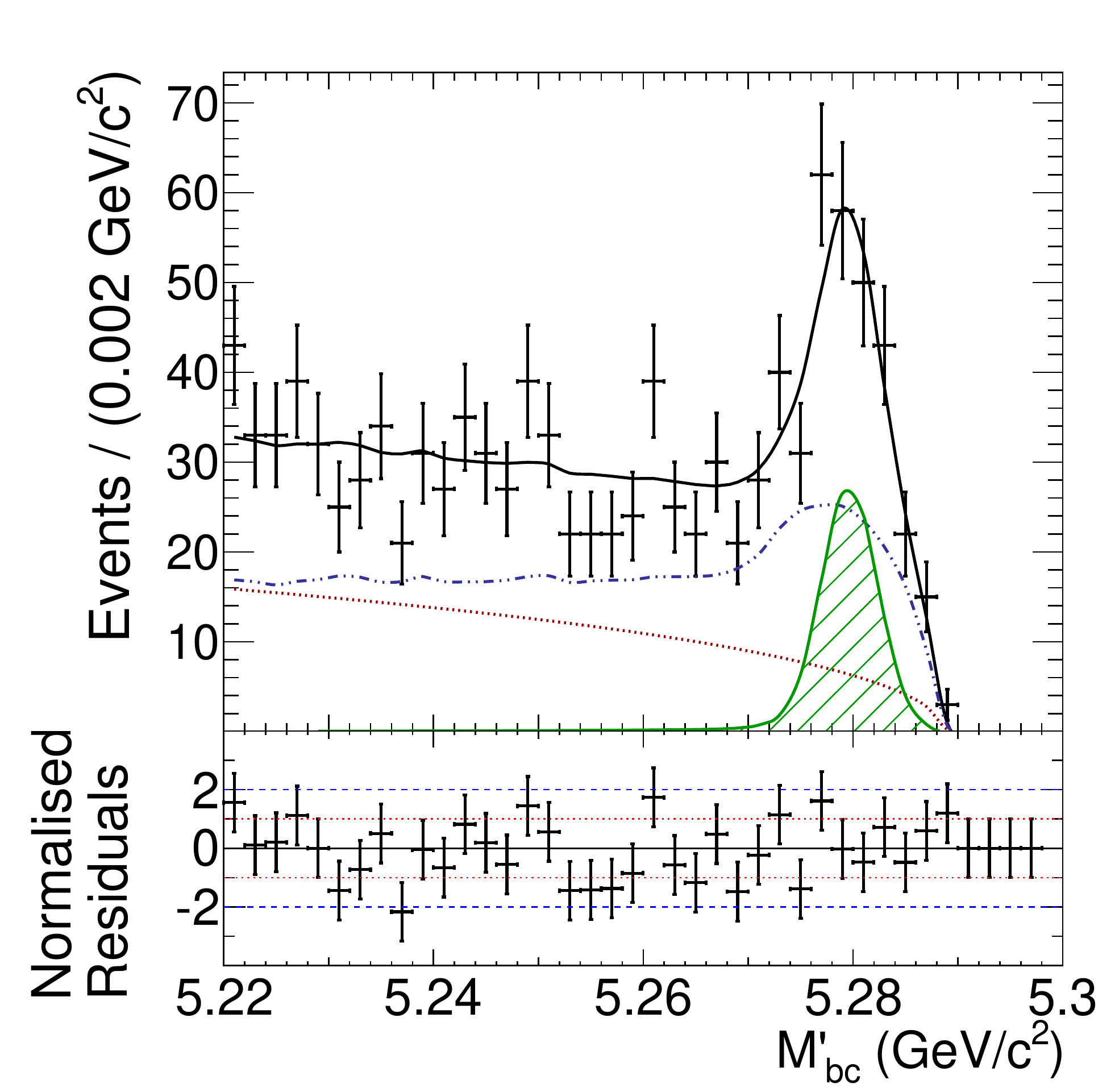}
 \includegraphics[width=0.49\columnwidth]{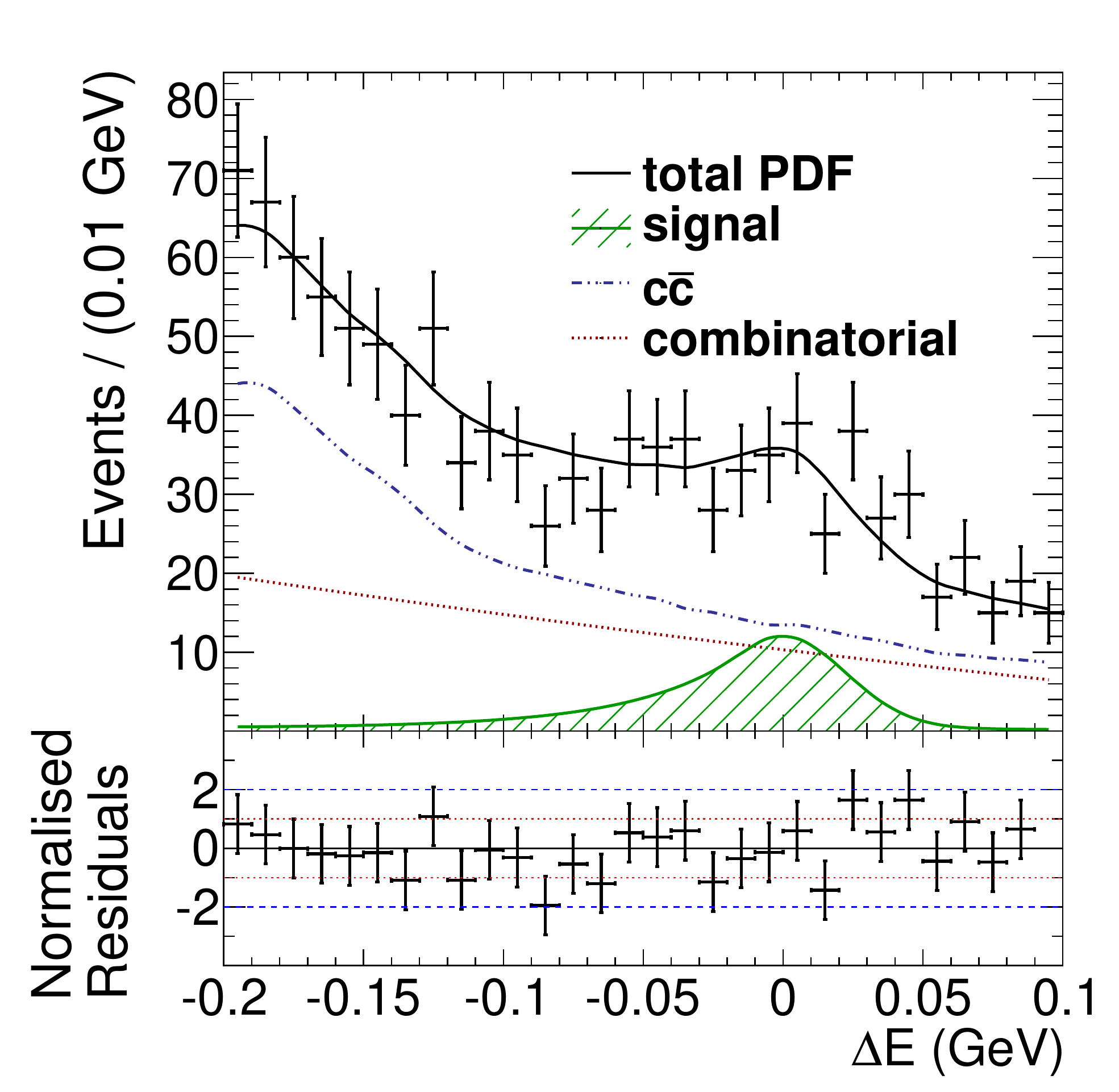}
 \caption{Projections of the fit to the \Bopp\ data in the entire fit region onto \mbc\ (left) and \de\ (right). Points with error bars represent the data and the solid 
 black curves represent the fit results. Green hatched curves show the \Bopp\ signal component, blue dash-dotted curves show the $c\bar{c}$ background 
 component, and red dotted curves indicate the combinatorial background.}
 \label{fig_data_p2p}
\end{figure}

From the fit to the data containing 1090 \Bopp\ candidates, we obtain the bias-corrected branching fraction
\begin{equation}
  {\cal B}(\Bopp) = (1.17 \pm 0.17)\times 10^{-5}.
  \label{p2p_bf_result}
\end{equation}
 The branching fraction corresponds to $85\pm12$ signal events, of which 
$38\pm8$ are leptonic and $47\pm9$ are hadronic, $628\pm65$ events originate from other $b\to (c\bar{c})q$ decays and 
$377\pm103$ events belong to the combinatorial background. All uncertainties here are statistical. Fit projections to the data are shown in Fig.~\ref{fig_data_p2p}.

Systematic uncertainties from various sources are considered. They are estimated with both model-specific and -independent studies and cross-checks. The ${\cal B}(\Bopp)$ 
systematic uncertainties are summarized in Table~\ref{tab_syst_p2p}. 

\begin{table}[h!]
  \caption{Systematic uncertainties of the \Bopp\ branching fraction.}
   \renewcommand{\arraystretch}{1.1}%
  \centering
  \begin{tabular}
    {@{\hspace{0.2cm}}l r@{\hspace{0.2cm}}}     
    \hline
    \hline
    Category 	& $\delta{\cal B}(\Pgyii\Pgpz)$ $[\%]$   \\
    \hline
    $N_{B\bar{B}}$ & 1.4 \\ 
    \Pgpz\ reconstruction & 4.0 \\
    ${\cal B}(\Pgyii\to\APlepton\Plepton)$ & 3.0 \\ 
    ${\cal B}(\Pgyii\to\PJpsi\Pgpp\Pgpm)$ & 0.5 \\ 
    ${\cal B}(\PJpsi\to\APlepton\Plepton)$ & 0.3 \\ 
    Electron ID & 0.7 \\ 
    Muon ID & 0.9 \\ 
    Hadron ID & 1.3\\ 
    Tracking & 1.7\\ 
    Misreconstruction & 0.3\\ 
    Parametric shape & 0.9\\ 
    Nonparametric shape & 1.4\\ 
    Peaking $b\to(c\bar{c})q$ background in \mbc & 1.7\\ 
    Peaking background in \mbc\ and \de & 2.2\\ 
    Correction factors & 0.9\\ 
    Fit bias & 0.6\\ 
    \hline	
    Total & 6.7\\ 
    \hline
    \hline
  \end{tabular}
  \label{tab_syst_p2p}
\end{table} 

The systematic uncertainty due to the error on the total number of $B\bar{B}$ pairs is calculated from the 
on- and off-resonance luminosity, taking into account the efficiency and luminosity scaling corrections~\cite{BBook}. The dominant systematic uncertainty arises 
from the \Pgpz\ reconstruction and is evaluated by comparing data-MC differences in the yield ratios between $\eta'\rightarrow\Pgpz\Pgpz\Pgpz$ and 
$\eta'\rightarrow\Pgpp\Pgpm\Pgpz$. We also consider the systematic uncertainties originating from the knowledge of the \Pgyii\ and \PJpsi\ decay branching fractions used 
to calculate the efficiency. 
We apply the percentage error on their world averages~\cite{PDG} as a systematic uncertainty. The electron and muon identification efficiency 
uncertainties were obtained from separate Belle studies of the two-photon processes $\Pep\Pem\to\Pep\Pem\APlepton\Plepton$ and of $\PJpsi\to\APlepton\Plepton$, where 
$\ell=e,\mu$. The uncertainty in the reconstruction efficiency due to the hadron identification is determined 
using $D^{*+}\rightarrow D^{0}[K^{-}\Pgpp]\Pgpp$ decays, where the hadron identity is unambiguously determined by its charge. The uncertainty due to 
the tracking efficiency is calculated by comparing data-MC differences in the reconstruction efficiencies of 
$D^{*\pm}\rightarrow D^{0}[K^{0}_{S}\{\Pgpp\Pgpm\}\Pgpp\Pgpm] \pi^{\pm}$. The hadron, electron and muon identification and tracking uncertainties are 
weighted by the reconstruction efficiencies of the corresponding $B$ decay modes.
The misreconstructed signal uncertainty is obtained by varying the misreconstructed fraction by 
$\pm20\%$ of its value, which is a conservative estimate.  The parametric and nonparametric shapes describing 
the background are varied within their uncertainties. For nonparametric shapes ({\it i.e.}, histograms), we modify the histogram PDFs bin by bin according to a 
Poisson distribution and extract the branching fraction from a fit to the data. We perform 300 tests with such modified histogram PDFs and take the 
width of the resulting Gaussian branching-fraction distribution as a systematic uncertainty. We find that the decay $B^{0}\to\Pgyii K^{0}_{S}[\pi^{0}\pi^{0}]$ peaks in the signal region 
of \mbc. The $B^{0}\to\Pgyii K^{0}_{S}[\pi^{0}\pi^{0}]$ yield in the $b\to (c \bar{c})q$ background sample is varied by the uncertainty of its world average branching fraction 
and the resulting difference in the \Bopp\ branching fraction is taken as a systematic uncertainty. The number of peaking background events obtained from the sideband study is 
varied by one standard deviation ($\sigma$), and the difference in the branching fraction is assigned as a systematic uncertainty.
The same approach is used for the \mbc\ and \de\ correction factors. Half the branching-fraction 
fit bias obtained from pseudo-experiments is taken as an additional systematic uncertainty. The total systematic uncertainty is 6.5\% of the \Bopp\ branching fraction.

\begin{figure}
 \centering
 \includegraphics[width=0.9\columnwidth]{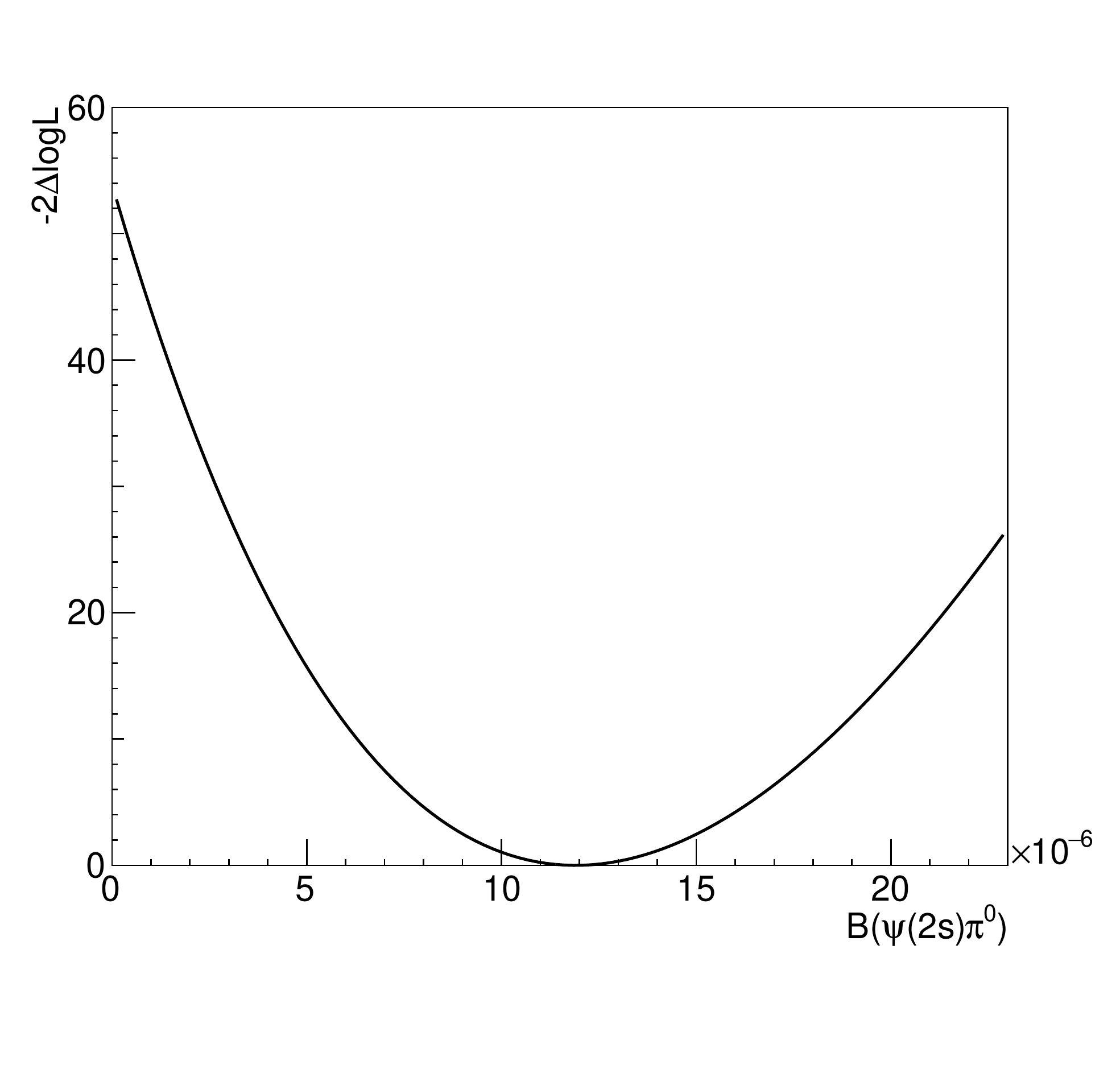}
 \caption{${\cal B}(\Bopp)$ likelihood scan. The likelihood is convolved with an additive systematic uncertainty.} 
 \label{bf_scan_p2p}
\end{figure}

We perform a likelihood scan to obtain the statistical significance of our branching fraction measurement. We convolve the  ${\cal L}$ distribution with a
Gaussian with a zero mean and a width equal to the systematic uncertainty. The change in the $-2\log{\cal L}$ distribution as a function of the branching 
fraction is shown in Fig.~\ref{bf_scan_p2p}. The statistical significance of $7.2\sigma$ is determined from 
$\sqrt{-2\Delta\log{\cal L}}$, where $\Delta \log{\cal L}$ is the likelihood difference between zero and the observed branching fraction.
This includes the systematic uncertainties.

In summary, we report a measurement of the \Bopp\ branching fraction based on the full Belle data set collected at the $\Upsilon(4S)$ resonance. We obtain 
${\cal B}(\Bopp) = (1.17\pm0.17\text{(stat)}\pm0.08\text{(syst)})\times10^{-5}$. Our results are consistent with the na\"{i}ve expectation that the \Bopp\ to 
$B^{0}\to \psi(2S)K^{0}_{S}$ branching fraction ratio should be similar to the $B^{0}\to J/\psi \pi^{0}$ to 
$B^{0}\to J/\psi K^{0}_{S}$  ratio. The ${\cal B}(\Bopp)$ result has a significance of $7.2\sigma$, which indicates the first observation of the decay \Bopp.

We thank the KEKB group for excellent operation of the
accelerator; the KEK cryogenics group for efficient solenoid
operations; and the KEK computer group, the NII, and 
PNNL/EMSL for valuable computing and SINET4 network support.  
We acknowledge support from MEXT, JSPS and Nagoya's TLPRC (Japan);
ARC (Australia); FWF (Austria); NSFC and CCEPP (China); 
MSMT (Czechia); CZF, DFG, and VS (Germany); DST (India); INFN (Italy); 
MOE, MSIP, NRF, BK21Plus, WCU and RSRI  (Korea);
MNiSW and NCN (Poland); MES and RFAAE (Russia); ARRS (Slovenia);
IKERBASQUE and UPV/EHU (Spain); 
SNSF (Switzerland); NSC and MOE (Taiwan); and DOE and NSF (USA).

\end{document}